\documentstyle{mn}

\begin{document}
\title{The structure of spiral galaxies - II. Near-infrared properties of spiral arms}
\author[M.S. Seigar \& P.A. James]{M.S. Seigar \& P.A. James\\
Astrophysics Group, Liverpool John Moores University, Byrom Street,\\
Liverpool, L3 3AF, U.K.\\
email: {\tt mss} \& {\tt paj} {\tt @staru1.livjm.ac.uk}}
\maketitle

\begin{abstract}

We have imaged a sample of 45 face-on spiral galaxies in the K-band, to
determine the morphology of the old stellar population, which dominates
the mass in the disk. The K-band images of the spiral galaxies have been
used to calculate different characteristics of the underlying density
perturbation such as arm strengths, profiles and cross-sections, and
spiral pitch angles.  Contrary to expectations, no correlation was found
between arm pitch angle and Hubble type, and combined with previous
results this leads us to conclude that the morphology of the old stellar
population bears little resemblance to the optical morphology used to
classify galaxies.

The arm properties of our galaxies seem inconsistent with predictions
from the simplest density wave theories, and some observations, such as
variations in pitch angle within galaxies, seem hard to reconcile even
with more complex modal theories. Bars have no detectable effect on arm
strengths for the present sample.

We have also obtained B-band images of three of the galaxies. For
these galaxies we have measured arm cross-sections and strengths,
to investigate the effects of disk density perturbations on star formation in
spiral disks. We find that B-band arms lead K-band arms and are narrower
than K-band arms, apparently supporting predictions made by the large
scale shock scenario, although the effects of dust on B-band images may
contribute towards these results.

\end{abstract}


\section{Introduction}

Ever since the pioneering work of Lin \& Shu (1964, 1966) many
theoretical models have been proposed to explain the existence of spiral
structure in disk galaxies. Many are based upon the Lin-Shu Hypothesis
(e.g. Lin, Yuan \& Shu 1969; Roberts 1969; Roberts, Roberts \& Shu 1975)
which describes spiral patterns as density waves, which cause compression
of the gas component as it flows through the arms leading to subsequent
star formation. Modal theories (Bertin et al. 1989a, 1989b; Bertin \& Lin
1996) are more complex density wave models, which describe disks of
galaxies as resonant cavities within which density waves of different
modes can co-exist and interfere to produce a range of observed
phenomena. Tidal models (Toomre
\& Toomre 1972; Kormendy \& Norman 1979) set up spiral structure through
transient density waves, which are caused by the tidal field of a nearby
neighbour. Bars can also potentially drive spiral structure (Sanders \&
Huntley 1976). In this case the formation of arms is driven by the effect
of the bar potential on the dissipative interstellar medium. Finally
Stochastic Self-propagating Star Formation (SSPSF) can describe
flocculent structure. Here, short irregular arms are produced due to
shearing of a part of the disk that has recently formed stars (Gerola \&
Seiden 1978). SSPSF alone cannot describe Grand-Design structure, but
some models (e.g. Sleath \& Alexander 1995, 1996) use SSPSF with a weak
imposed density wave to describe such global spiral modes.

Many of the recent advances in this area have arisen from a study of the
atomic and molecular gas component in spiral galaxies, through studies of
HI and CO line emission.  This has permitted the detailed mapping of both
the distribution and velocity field of the gas in the spiral arms of
nearby galaxies, e.g. M81 (Visser 1980) 
M51 (Tilanus \& Allen 1991, 1993; Rand 1993; Nakai
et al. 1994), M100 (Knapen 1993; Rand 1995), NGC~3627 (Reuter et al.
1996) and NGC~6946 (Regan \& Vogel 1995).  These studies find streaming
velocities of gas through the spiral arms of order a few 10s of
km~s$^{-1}$, and also find offsets between the peaks of the gas density
and the old stellar population in the arms and the star formation as
revealed by H$\alpha$ emission.  All of these findings are in general
agreement with the predictions of density wave theories, with the gas
being compressed and shocked as it flows into the spiral arm, and
subsequently forming stars downstream from the density peak of the spiral
arm. However, these studies are only possible in the strongest arms of
the nearest spiral galaxies, and hence yield little information on the
global importance of density waves in the general population of
disk galaxies. Here we undertake the complementary approach of
calculating the strength of density waves as mapped out by the old
stellar population of a large sample of galaxies, using near-IR K-band
($2.2\mu m$) imaging. In this waveband the old stellar population is
observed (Rix \& Rieke 1993), making this the best observational tracer
of the stellar mass distribution. Also, K--band measurements are less 
affected by extinction than measurements in the optical.

In a previous paper (Seigar \& James 1998 - hereafter Paper I) we
described the properties of the bulges, disks and bars of a sample of 45
galaxies. In this paper we describe how spiral arm structures can be
extracted for these galaxies and determine quantitative parameters for
them. In section 2 we describe the observations; section 3 describes the
data reduction and extraction of all of the spiral arm parameters used in
the analysis; section 4 then draws on these parameters in discussing, in
turn, each of the theoretical models of spiral structure; in section 5 we
describe tests of models of star formation in spiral galaxy disks and in
section 6 we summarise our findings.

\section{Observations}

A sample of 45 face-on spiral galaxies, selected 
using the NASA/IPAC Extragalactic Database (NED)\footnote[1]{The NASA/IPAC 
Extragalactic
Database (NED) is operated by the Jet Propulsion Laboratory, California
Institute of Technology, under contract with the National Aeronautics and
Space Administration.},
was observed using the
infrared camera IRCAM3 on the United Kingdom Infrared Telescope (UKIRT).
IRCAM3 utilises a 256x256 array with a pixel size of 0.286 arcsec, and
the observations presented here were made in the J and K bands. The
observing dates were 1995 February 5 \& 6, 1995 November 18, 19
\& 20 and 1996 March 22 \& 23. Each galaxy was observed for a total
integration time of 1800 seconds in the K-band, and 540 seconds in the
J-band.  All measurements presented here are taken from the K band images
unless otherwise stated. For flat-fielding purposes, sky frames were
obtained using equal integration times on areas of nearby blank sky. A
dark frame with the same on-chip exposure time was taken both for the
galaxy frames and the sky frames.  Standard stars, taken from the UKIRT
faint standards list, were observed throughout all nights to provide
photometric calibration.

This sample spans a range of Hubble types classified as Sa to Sd, 
including barred and non-barred galaxies as from the Third Reference Catalogue
of Bright Galaxy (de Vaucouleurs et al 1991; hereafter RC3). 
The details of the sample are explained in paper
I. The Hubble types of the galaxies are listed in column 2 of table 1.

\section{Extraction of Spiral Arm Structures}

In this section we describe the technical details of the extraction of
quantitative descriptors of spiral structure in the 45 spiral galaxies.
These parameters will then be used in subsequent sections to test
theories of spiral structure.

\subsection{Ellipse Fitting}

The RGASP package was used to fit ellipses to the isophotes in the
reduced galaxy images. The ellipticity of the outermost ellipses was then
adopted and the PROF routine, which takes medians around ellipses at increasing
radii, was run keeping this ellipticity and position angle constant. 
The REBUILD facility in RGASP enables a simulated
galaxy to be made up using the ellipticity and position angle and the
median surface brightness calculated around these ellipses, and
subtracting this rebuilt image from the original image removes the
elliptically--symmetric bulge and disk components.  This procedure leaves
an image of the bar and spiral arms.  Plotting the surface brightness of
the fitted ellipses as a function of semi--major axis gives light--profiles, 
which were used to do a one-dimensional
bulge-disk decomposition (see paper I).

\subsection{Deprojection}

The images were next deprojected, to remove the effects of galaxy
inclination on the derived bar and arm parameters. Each
galaxy was rotated so that its major axis was parallel with the y-axis of
the Cartesian coordinate frame. The minor axis of the galaxy was then
stretched along the x-axis by a factor determined by the ellipticity of
the galaxy, resulting in the minor axis becoming equal to the major axis.
The galaxies in the sample are either face-on or nearly face-on, so
deprojection effects are small (typically the ratio of the minor-axis
to the major-axis, $b/a\simeq 0.8$).

\subsection{Conversion to Polar Coordinates}

Each of the deprojected difference images was converted from Cartesian to
polar coordinates using a method similar to that of Elmegreen, Elmegreen
\& Montenegro (1992). The Cartesian image was mapped onto the new polar
grid by calculating the Cartesian coordinates for each of the polar
coordinate pixels. The number of counts associated with the calculated
x-y position was then put in that polar coordinate position. The polar
frame was divided into 540 steps of $1^{\circ}$ in the $\theta$ direction 
(i.e. wrapping around the galaxy one and a half times), and into 600 
in the radial direction where 1 division was
equal to $0.01\ln{(r)}$ and $r$ is measured in pixels.  This gave us
sufficient oversampling to avoid loss of resolution even in the outer
parts of the polar grid, and displaying one and a half turns makes it
easier to identify spiral arms as continuous features.

\subsection{Measurement of Spiral Pitch Angle}

Spiral structure is assumed to have the form of a logarithmic spiral, given by
\begin{equation}
n\theta=\Lambda \ln{\left(\frac{r}{r_{0}}\right)}
\end{equation}
where $n$ is the number of arms, $\theta$ is the azimuthal angle,
$\Lambda$ is a constant which defines the winding angle, $r$ is radius
and $r_{0}$ is a scale length.

The pitch angle $i$ is given by
\begin{equation}
\cot{(i)}=r\left|\frac{d\theta}{dr}\right|.
\end{equation}

It can therefore be seen that
\begin{equation}
\cot{(i)}=\frac{\Lambda}{n}=\frac{d\theta}{d(\ln{(r)})}
\end{equation}
i.e. the gradient of the arm in a $\theta$ vs $\ln{(r)}$ plot. This gradient
was calculated by clicking along the ridge of the spiral arm with the cursor
and fitting a straight line to these points. 
If the pitch angle of the arm changed with radius then we calculated an
average pitch angle for that arm. Where galaxies had multiple arms the average
pitch angle over all the arms was calculated. The errors in the pitch angle
are calculated from the errors in the straight line fit, which is calculated
using the residuals between the line and the data points. This is therefore
dominated by the change in pitch angle with radius, in galaxies where this 
change is significant.
The calculated pitch angles for all of the galaxies in this sample are 
listed in column 3 of table 1.

\subsection{Measurement of Arm Strength}

\subsubsection{Definition of Equivalent Angle}

We have defined a new quantity, Equivalent Angle (EA), which parametrises
the strength of arms, and was also used for bars in paper I. It is
defined as the angle subtended by a sector of the disk which contains the
same amount of flux as does the spiral arm, within the same radial
limits. The advantage of using equivalent angle is that it is
unaffected by seeing and other resolution effects unlike, for example,
arm-interarm contrast (see paper I for a more detailed discussion).

\subsubsection{Calculation of Equivalent Angle}

The polar coordinate images were used for the calculation of EA of spiral
arms. Using the gradient of the arms (calculated whilst working out the
pitch angle) the arms were transformed into horizontal features, by
shearing columns of pixels according to the calculated gradient. Although
this is not necessary here, it is necessary for calculating arm 
cross-sections (see section 3.6) and it also makes the calculation of arm
EA easier. The arm
was then divided into strips, each covering a small range in $\ln{(r)}$, and
the equivalent angle was calculated for each strip, by dividing the arm
flux by the flux from the rebuild image in the corresponding strip. In
this way it was possible to calculate the equivalent angle as a function
of radius for all of the galaxies in the sample. 

The overall EA was calculated by taking the mean value of EA over the
entire detectable range of the arm. The cutoff radii for this calculation
are taken where the EA falls below $3^{\circ}$. The overall EA values are
listed in column 4 of table 1.

\subsubsection{Arm Profiles}

Figure 1 shows how arm EA varies with radius for four typical cases, NGC
1219, NGC 3512, UGC 11524 and IC 357. The start and end points are
defined as in the previous section, except in the case of IC 357 where
the starting point is at the end of the bar. 

\subsection{Calculation of Arm Cross--sections}

Using the plots described in the previous section, where the arms are
transformed into horizontal features, it is simple to calculate profiles
across spiral arms by summing the flux in rows, i.e. along the arms.
Figure 2 shows two such cross-sectional profiles (top row) for
representative galaxies, with their decompositions into symmetric (middle
row) and antisymmetric (bottom row) parts. Also shown in figure 2 is a 
540$^{\circ}$ cut of IC 1809. The profiles can also be used to
derive FWHM arm widths, measured in degrees of azimuthal angle, which are
listed in column 6 of table 1. The errors on these FWHM depend on the errors
on the measured gradients (i.e. pitch angles). With the error typically
$<$5\% on the pitch angle it is found that the error on the FWHM is 
typically $<$10\%. This error has the effect of 
moving the peak of the cross-section
by up to 10\% and broadening the arm cross-section by 10\% and this leads
to an error in the calculated symmetries of $\pm$5\%. 
Dividing these widths into the
Equivalent Angle from column 4 of table 1 gives a good indication of the
arm contrast, or degree of non-linearity, averaged along the arm.
This is because EA is a measure of the flux in the arm with respect to the 
disk and the arm cross-section FWHM is a measure of the width of the arm in 
which this is contained. The shock strength is dependent upon the arm strength
and we also expect stronger shocks in narrower arms. Therefore the ratio of
arm EA to arm FWHM is a plausible indicator of the relative shock strengths
in spiral arms.

The decomposition into symmetric and antisymmetric parts was performed using
\begin{equation}
I(r,\theta)=\frac{I(r,\theta)+I(r,2\phi-\theta)}{2}+\frac{I(r,\theta)-I(r,2\phi-\theta)}{2}
\end{equation}
where the first and second terms represent the symmetric and antisymmetric
parts, respectively, with respect to the position angle $\phi$ of the major 
axis of the arm when plotted in $\theta$ vs $\ln{(r)}$ coordinates (Ohta,
Hamabe \& Wakamatsu 1990). The value of $\phi$ was determined as being the 
data point that had the highest surface brightness. If there were two data 
points near the peak with similar surface brightness, then the value of $\phi$ 
was taken to be half-way between these two points. When the value of $\phi$ is 
changed by up to 2 pixels (the typical error in determining the value of 
$\phi$) the asymmetries change by $<$2\%.

\subsection{Fourier Analysis}

\subsubsection{Method}

We have measured the dominant modes of spiral structure in these galaxies
using a simple Fourier technique adapted from the method used by
Gonz\'alez \& Graham (1996). Three 10--pixel wide strips were taken along
the $\ln{(r)}$ coordinate of the $\theta$ vs $\ln{(r)}$ frame. These were
binned up to create a one-pixel-wide strip suitable for analysing with a
1-D Fast Fourier Transform (FFT) routine. The effect of smearing of the
arms for a 10 pixel wide strip is small. An FFT of each of the strips was
then taken, resulting in three power spectra, which were summed to increase
the signal-to-noise ratio. As the FFT is taken over $360^{\circ}$, the
power in all modes up to $m=360$ is calculated, although we are only
interested in modes up to $m\simeq 6$. We average all the modes from
$m=1$ to $m=6$ and quote the power of individual modes as a ratio with 
this average, in order to overcome any systematic noise effects (see
section 3.7.2). The strongest mode for each galaxy in the sample is
listed in column 6 of table 1. The relative strengths of all the modes are
listed in table 2.

Figure 3a shows a histogram of the occurrence of the strongest mode in
our sample with a subset in black indicating this for galaxies with near
neighbours (see section 4.4 for a definition of near neighbours).
Figure 3b shows the same histogram with a subset for the 
strongly barred galaxies, where
a strongly barred galaxy is defined as a galaxy with a bar overall
equivalent angle greater than the average for the whole sample (average
bar EA=19.7$^{\circ}$).  These histograms will be discussed in section 4.

\subsubsection{Discussion of Noise Effects}

Fourier analysis can be subject to effects from noise, which may alter
relative mode strengths (Gonz\'alez
\& Graham 1996). We have investigated this problem in two ways, the first
of which was to add a simulated noise frame to a $\theta$ vs $\ln{(r)}$ galaxy
image, thereby doubling the pixel-to-pixel noise. Here, the image of
IC 1764 was used. We found that the mode 
strengths were changed by $<5\%$ even under this significant noise increase.

For the second test, we simulated a strip of constant radius across a 
$\theta$ vs $\ln{(r)}$ image
by adding together sine waves of different frequencies and introducing noise
to this model. The input mode strengths were typical of those of our
images. Even for models with pixel-to-pixel noise greater 
than our noisiest image, the FFT still finds the low order modes to have 
approximately the same power as was input.
Thus it can be assumed that any noise has a small effect on the Fourier
parameters presented here.

\section{Testing Models of Spiral Structure}

Having described the determination of all the necessary parameters from
the near-IR images, we will now investigate the implications for
various models of the formation of spiral structure in galaxy disks.
This section is organised with each mechanism being discussed in turn, to
see whether the new measurements support the mechanism as a strong contributor 
to the spiral structure.  It should be noted that the various mechanisms
discussed are by no means mutually exclusive, and it may be that several
or all of them contribute at some level to the structures seen in spiral
galaxy disks. 

\subsection{The Lin--Shu Hypothesis}

\subsubsection{Spiral Pitch Angle}

A fundamental prediction of the Lin--Shu hypothesis, and a  central feature
of the Hubble classification scheme, is that galaxies with greater mass
concentration in their central bulges should have more tightly wound
spiral structure.  This can be demonstrated by the following
back-of-envelope calculation.  Lin \& Shu (1964) derive the following
expression for the locus of the spiral pattern,
\begin{equation}
n(\theta-\theta_{0})=-\int_{r_{0}}^{r}\frac{[\kappa^{2}+\omega_{i}^{2}+(\omega_{r}-n\Omega)^{2}]}{(2\pi G\mu_{0})}dr
\end{equation}
where $\theta$ is the azimuthal angle, $\kappa$ is the epicyclic frequency,
$\mu_{0}$ is the fraction of mass in the disk, $\Omega$ is the pattern speed 
and $\omega_{r}$ and $\omega_{i}$ are the real and imaginary parts, 
respectively, of the material speed.

Using equations 5 and 2,
\begin{equation}
cot(i)=-\frac{1}{n}\frac{[\kappa^{2}+\omega_{i}^{2}+(\omega_{r}-n\Omega)^{2}]}{2\pi G\mu_{0}}
\end{equation}

Equation 6 contains the desired result, that galaxies with more dominant
disks should have spiral structure that is more loosely wound. Roberts et
al. (1975) converted this into a correlation between pitch angle and Hubble
type, which they assumed to be essentially determined by bulge-to-disk
ratio. They calculated the theoretical winding angle from the
bulge-to-disk ratios of a sample of galaxies and plotted this versus
Hubble type. The line drawn in figure 5 is their predicted correlation.
Roberts et al. (1975) also state that although the pitch angle has some
dependence on the dynamical properties of the disk, e.g. $\kappa,
\omega_i$ and $\omega_r$ in equation 6 above, its dependence upon the fraction of mass
in the disk is stronger. The dynamical properties, especially the
material speed, seem to correlate more strongly with the luminosity class
(van den Bergh 1960) of the galaxy than with Hubble type. 

Kennicutt (1981) found only weak correlations between pitch angle and Hubble
type, and between pitch angle and bulge-to-disk ratio. He also found a
correlation between pitch angle and the maximum rotational velocities
of spiral galaxies. This seems inconsistent with spiral density wave
theories as it shows that arm strength reflects the effects of differential
rotation. Kennicutt (1982) found that arm width increases smoothly with galaxy 
luminosity, with a slope consistent with broadening by galactic rotation (i.e.
galaxies with larger rotation speeds have broader arms). This
is consistent with the luminosity class (van den Bergh 1960) being dependent 
upon dynamical properties of galaxies (Roberts et al. 1975). Kennicutt \&
Hodge (1982) measured pitch angles of 17 galaxies from the sample of Roberts
et al. (1975) using H$\alpha$ data. They found that their measured pitch 
angles correlated well with the Roberts et al. (1975) model pitch angles, but 
that the model systematically underestimated all the pitch angles. These 
authors' work was performed using B--band and H$\alpha$ data which is strongly
affected by young stars, and so we have performed this using K--band data, 
which is known to have a significantly different morphology from optical
wavebands (e.g. Block et al. 1994).

We can directly test the important prediction of a correlation between
mass concentration and winding angle using our images. The pitch angle
(equation 2) quantifies how tightly wound spiral structure is, where
small pitch angles imply tightly wound structure. To parametrise the
central mass concentration we used the fraction of K-band light in the
disk ($L_{disk}/(L_{disk}+L_{bulge})$), calculated by numerical
integration over the profiles, from the galaxy centre to beyond the edge
of the observed galaxy, for both bulge and disk. This parameter is listed
in column 7 of table 1. (We use this parameter rather than the more
widely used bulge-to-disk ratio for consistency with Lin \& Shu 1964.)

A plot of pitch angle versus the fraction of light in the disk (shown in
figure 4) shows no correlation of the type predicted by Lin \& Shu
(1964). Figure 5 shows a plot of pitch angle against Hubble type, which
also lacks any correlation and is in very clear disagreement with the
line derived from the predictions of Roberts et al. (1975). 

There is an even simpler test which demonstrates that spiral arms cannot
have the fixed form indicated by equation 6. The lack of any
$\theta$-dependence predicts that the arms in any spiral galaxy should
have the same pitch angle at any given radius, in the Lin-Shu model. This
can be tested by looking at $\theta$ vs $\ln{(r)}$ images of galaxies, where
the pitch angle can be determined directly from the gradient of the arms.
This test is completely independent of the radial variation of dynamical
properties. Figure 6 shows a contour plot of the galaxy NGC 2503,
demonstrating that the two arms have very different gradients, and
therefore pitch angles, at the same radius. This occurs in 20
out of the 45 galaxies in our sample, and is also seen clearly in the
$\theta$ vs $\ln{(r)}$ image of M99 presented by Gonz\'alez
\& Graham (1996). Of these 20 galaxies, 9 have strong m=1 modes, i.e. a
degree of lopsidedness, which can produce variations in pitch angle.
 This is seen in galaxies which have a bulge which is 
offset from the centre of the disk as seen in UGC 3900 (see figure 8). 
Both this and the fact that pitch angle does not
correlate with Hubble type or B/D ratio seem to suggest that the simplest
form of density wave theory is at least incomplete. Indeed, it is not
clear that a superposition of spiral waves, as postulated by modal
theories, can produce this type of azimuthal variation in arm properties,
which is most easily reconciled with SSPSF models.

We find the lack of any correlation between
spiral winding angle and Hubble type quite surprising.  In paper I we
found that Hubble type correlates only weakly with bulge-to-disk ratio
(B/D), and we now find no correlation with arm morphology, at least in
the near-IR, raising the question of what does determine Hubble type.  
Other recent studies have found that the bulge surface brightness 
correlates strongly with Hubble type (de Jong 1996b) suggesting that only
bulge parameters are important in determining Hubble type.

Another possibility is that the mass of cold gas plays a key role
in determining galaxy type.  This is consistent with arguments of Block
\& Wainscoat (1991) and Block et al. (1994) who find that the Population
I disk of young stars, which dominates the optical appearance, can be
largely decoupled from the distribution of older stars. This can
explain the discrepancy between our findings and the optical study
of Kennicutt (1981). (Note, however,
that in section 5.1 we will report reasonably good agreement between arm
morphologies for 3 of our galaxies imaged in both the B and K bands, so
this decoupling cannot be complete.)

\subsubsection{Arm Cross-Sections}

We now consider cross--sections through spiral arms illustrated for some
of the more strongly--armed galaxies in figure 2, and the implications of
these for density wave theories.

Figure 2 shows that arm cross-sections for strong arms
are typically asymmetric at only
the 5-10\% level, which is consistent with density wave theory, assuming
that the K-band light traces the stellar mass density and is largely
unaffected by extinction and the shocks which perturb the gas density.
However, the Lin--Shu hypothesis predicts that density waves are
sinusoidal in nature with a $180^{\circ}$ period, leading to a FWHM of
$60^{\circ}$ for each arm of a 2--armed spiral (c.f. $120^{\circ}$ for a
one-armed mode, $40^{\circ}$ for 3--armed, $30^{\circ}$ for 4--armed etc).
The measured FWHM of the arm cross--sections vary between $10^{\circ}$ and
$70^{\circ}$ with typical values in the range $20^{\circ}$--$40^{\circ}$
(see column 6 of table 1). Thus the arms are almost always narrower than
expected for grand--design 2--armed structure, and in most cases $\sim$25\%
narrower than expected on the basis of the dominant Fourier mode of the
galaxy. The typical error on the FWHM is $<$10\% and the typical error
on the asymmetries is 5\%.
The 540$^{\circ}$ cut of IC 1809 also demonstrates that arm 
cross-sections are not sinusoidal. 
There are two possible explanations for this, the first of which is
that the Lin--Shu hypothesis may be oversimplistic (see section 4.2). 
The other explanation
may be that the arms are narrower due to the effect of young super-giant 
stars forming near the peak of the density wave. Rix \& Rieke (1993) showed 
that the affect of supergiants is small for M51, but Rhoads (1998) has shown
that supergiant stars contribute up to 30\% of the local luminosities, which
may have the effect of making arms appear somewhat narrower than predicted by
the Lin-Shu hypothesis.

Figure 15 shows K-band arm cross-sections for the galaxies for which we
have B-band data. These cross-sections appear less symmetrical than those
shown in figure 2. However, the galaxies in figure 2 have strong spiral
structure, whereas those in figure 15 have arms that are weaker and the
profiles are thus somewhat noisier. The asymmetries for these less
strongly-armed spirals are at about the 25\% level.

\subsection{Modal Theories}

\subsubsection{Fourier Analysis}

The narrowness of the arm cross-sections (shown in figure 2 and
described above) can be explained by modal theory (Bertin \& Lin 1996;
Bertin et al. 1989a, 1989b), where higher order modes can be superposed
on the $m=2$ mode resulting in narrower arms. The presence of such
multiple modes can be tested using Fourier decomposition of the disk
structure. Although low-order modes are expected to be dominant, high-order 
modes should also be present but with weaker amplitudes.

It can be seen from figure 3a that the most common dominant modes in
the sample are $m=1$ and $m=2$, each of which dominates the disk
structure of about one-third of the galaxies, whilst the $m=4$ mode
accounts for most of the remainder (a quarter of the total). The $m=3$
mode dominates in only 4 out of 45 cases, and the relative weakness of
the $m=3$ mode is probably the strongest result of this analysis. 

Rix \& Zaritsky (1995) performed a Fourier analysis of 18 face-on
spirals, imaged in the K band. They found that about half of the galaxies
in their sample had a strong two-armed spiral component with an
arm-interarm contrast of about 2:1 and consequently strong $m=2$ and
$m=4$ modes. Gonz\'alez \& Graham (1996) performed a detailed study of
the galaxies M99 and M51 including a Fourier analysis of both galaxies.
They looked at the dependence of $m=1$ to $m=6$ modes with radius and
found that the $m=2$ mode was the strongest at all radii in the
grand-design spiral M51, and that M99 was dominated by the $m=2$ mode at
most radii. This was found to be true in both optical and K-band images
of M99. Analysis of a sample of seven galaxies in the K$^{\prime}$-band
($2.1\mu m$) reported by Block et al. (1994) finds structure dominated by
the $m=1$ and $m=2$ modes, with the lack of higher modes being attributed
to damping of such modes in the old stellar disk by the Inner Lindblad
Resonance (ILR). 

We can therefore conclude that the frequent occurrence of the low order $m=1$
and $m=2$ modes is in agreement with predictions from modal theories, as
is the suppression of the $m=3$ mode. The dominance of the $m=4$ mode in 9 
out of 45 cases, however, cannot be explained by modal theories, which
predict that such modes should be damped by the ILR.

\subsubsection{Arm Profiles}

The arm profiles of the galaxies NGC 1219 and NGC 3512 (shown in figure
1) show some evidence for modulation (Bertin \& Lin 1996; Bertin et al.
1989a, 1989b), i.e. peaks and troughs in the radial light profiles of
arms of spiral galaxies. These authors interpret modulation as the result
of interference between different wave packets within the disk of the
galaxy. Our measurement of the same effect in K-band images demonstrates
that this is not simply due to extinction, but represents a true
modulation in arm strength.  Of the 16 galaxies with sufficiently long
and strong arms to test clearly for this effect, 4 show clear
modulation and a further 6 show some indication of such effects.

These modulation effects are superimposed on arm profiles generally
showing an increase in EA at small radii, which then turns over and
decreases. This implies that in the inner disk, the scale length of the
arms is greater than that of the disk (i.e. the arm surface brightness
falls off less rapidly than the disk surface brightness) and in the outer
disk the scale lengths of arms are smaller than disk scale lengths. This
is in agreement with the result of Gonz\'alez \& Graham (1996) who found
a peak in the arm-interarm contrast at intermediate radii.

\subsection{Driving by Bars}

\subsubsection{Testing models of bar-driven spiral structure}

Bars may be responsible for the driving of spiral structure (Sanders \&
Huntley 1976) but Sellwood \& Sparke (1988) have shown that that bar
driving is only important for the most strongly-barred spiral galaxies.
In addition, it is thought that bar forcing should not extend far outside
the region where the bar potential is strong (Sanders
\& Tubbs 1980).

We have tested models of bar-driven spiral structure by looking for a
correlation between bar strength and arm strength. The result of this can
be seen in figure 7, which  shows a plot of arm strength in the inner part of
the disk against bar strength. There is no correlation. It would
therefore appear that bars do not strongly affect the strength of spiral
arms, even in the central regions of disks.

\subsubsection{Results of the Fourier Analysis}

Figure 3b shows enhancement of the $m=3$ and $m=4$ modes for galaxies
that we have classed as strongly barred, relative to the sample as a
whole. A Kolmogorov-Smirnov (KS) test showed that this result is not
significant. If this result is confirmed by analysing larger samples of
galaxies in the same way, then it would be surprising as bar-driven models
predict strong 2-armed spiral structure (e.g. Sanders \& Huntley 1976; 
Sanders 1977).

\subsubsection{Relation Between Arm Morphology and Bars}

The relation between the morphology of bars and arms has been discussed
by Sellwood \& Sparke (1988), who claim that bars and arm structures may
have different pattern speeds. If this were the case then galaxies would
sometimes have arms starting ahead of the bar ends, although their
simulations show that this should not happen very often. They find that
in 100 time steps of their model there are two steps for which the ends
of bars and the beginnings of arms do not meet. We have observed this
behaviour in 2 out of 15 strongly barred galaxies. Our image of UGC~3900
(figure 8) shows quite evidently that its arms do not start at the end of
the bar, and IC~568 shows a similar morphology. The qualitative agreement
between the appearance of these galaxies and the simulations is very
good, but it may happen somewhat more frequently than predicted by
Sellwood \& Sparke (1988).

\subsection{Testing models of spiral structure involving tidal interactions}

Kormendy \& Norman (1979) proposed an alternative explanation to spiral
density waves for the maintenance of global spiral structure.  Noting
that the `spiral winding problem' is only actually a problem in the
differentially rotating parts of disks, they postulated that grand-design
structure may only occur in one of two situations. Firstly, some
galaxies may have grand-design spiral structure only within the radius
within which solid body rotation occurs, and thus completely avoid the
`spiral winding problem'. Kormendy and Norman then claim that galaxies
exhibiting grand-design spiral structure outside this radius could be
undergoing tidal interactions with other nearby galaxies, or be strongly
barred. In these cases the driving field from the neighbour or bar
maintains spiral structure even in the regions of the disk subject to
differential rotation (although note our conclusion in section 4.3.1).
Studying a sample of 54 galaxies, they found that 45 exhibited global
spiral patterns, and of these, only two clearly show global spiral
structures in the differentially-rotating part of the disk, with no bars
or neighbours to provide a driving field.  All others fitted into the
proposed model.

Observational support for this view has been provided by Elmegreen,
Seiden \& Elmegreen (1989) who observed that, generally, grand-design
spiral galaxies are found in richer environments than flocculent spirals
(i.e. grand-design spiral structure galaxies are more likely to have
undergone a tidal interaction with another galaxy in their history).  We
now test these results with our data.

Kormendy \& Norman (1979) define a tidally-significant neighbour galaxy
as one within 3--5 galaxy diameters and 1 magnitude (V-band) 
of the spiral galaxy 
in question. We have relaxed this somewhat to 6 diameters because few of 
the galaxies in
our sample meet the more stringent criterion, which is, in any case,
somewhat arbitrary, particularly since there is time for galaxies to move
apart between the time of interaction and the development of spiral
structure. We performed a search for near neighbours using the Digitised
Sky Survey and diameters (B$_{25}$) from the RC3. The apparent visual 
magnitudes of the near neighbours are comparable to those of the 
sample galaxies.

Figure 9 shows a histogram of the distribution of arm strengths (in
terms of equivalent angle) in the whole sample with the galaxies with near 
neighbours indicated in black. It can be seen that 4 out of 7 galaxies
with arm equivalent angle greater than $35^{\circ}$ have near neighbours
whereas only 12 out of 45 galaxies in the sample have companions. Whilst
a KS test shows that this difference is not statistically
significant due to the small
numbers involved, there is some indication here of a relation between arm
strength and tidal interactions. This conclusion is strengthened by
figure 3a which shows that the presence of near neighbours enhances
the $m=2$ mode, with this being the dominant mode in 7 out of 12 galaxies
with nearby neighbours, compared to only 7 out of 33
galaxies with no apparent neighbours.

Figure 10 shows a histogram of the distribution of the total
azimuthal angle over which the strongest arm of each galaxy in our sample 
can be detected, both for the whole sample and the galaxies with
near neighbours (black). This figure shows no evidence of near
neighbours enhancing global modes which would be represented by
longer spiral arms subtending larger azimuthal angles.

It therefore seems that the presence of nearby galaxies may affect
the strength of arms and enhance the $m=2$ mode, but they have no effect
on the angular extent over which arms can be detected.  This provides
some support for the model presented by Kormendy and Norman (1979), but
they also require driving by bars to be efficient at generating and
maintaining global arm structures, and we find no evidence for this
effect (section 4.3.1).

\subsection{Arm Strengths as a Function of the Underlying Galaxy Properties}

Kormendy \& Norman (1979) claim that a dominant bulge will prevent the
formation of global spiral arms in galaxies due to the 
differential rotation caused by a strong central mass concentration. We
can test this by determining the azimuthal angle that arms subtend as a
function of bulge-to-disk ratio.  This is plotted in figure 11, which
shows a hint of a deficit of global arm structures for galaxies with
dominant bulges, i.e. no points lying in the upper right-hand corner of
the plot. This is consistent with the predictions of
Kormendy \& Norman (1979).

Toomre (1964) claimed that galaxies with heavier disks are more unstable to
perturbations. Galaxies with heavier disks should have brighter disk central 
surface brightnesses and if these `heavy disks' were more unstable we would 
expect to obtain larger arm equivalent angles for them. Figure 12 shows a 
plot of arm equivalent angle against disk central surface brightness. No
correlation is shown but there does seem to be a lack of 
low-surface-brightness galaxies with large arm EA.

\section{Relation Between Arm Strength and Star Formation}

In this section we try to distinguish between the two main theories for
the occurrence of star-formation in spiral galaxy disks. The first of these
is the Large Scale Shock Scenario (Roberts 1969), which predicts that as
gas moves through the density wave it becomes shocked and star formation
results.  In this case star formation rate depends on the strength of the
shock and thus should correlate with measured arm strength.  However,
Elmegreen \& Elmegreen (1986) claim that density waves cannot trigger star
formation because their contrast above the disk is not high enough. They
suggest that although density waves are responsible for organising
material in spiral arms, some other mechanism must be responsible for
forming stars. Sleath \& Alexander (1995, 1996) suggest that this
mechanism is Stochastic Self Propagating Star Formation (Gerola \& Seiden
1978).  In this type of model, no strong correlation would be expected
between arm strength and star formation rate in the disk.

We have used the IRAS $60\mu m$ and $100\mu m$ fluxes
to calculate the far-infrared
luminosity for the galaxies in our sample for which IRAS data
was available. The far-infrared luminosity can be used as a
quantitative indicator of star-formation rates (e.g. Spinoglio et al.
1995). This was divided by the K-band luminosity of the galaxies
to compensate for differences in their overall size,
and then compared with their overall arm strengths in EA normalised to the
arm FHWM to see if shocks in the arms
are triggering the formation of stars. The far-infrared
luminosity in terms of the $60\mu m$ and $100\mu m$ flux 
is given by Lonsdale et al. (1985) as
\begin{equation}
L_{FIR}=3.75 \times 10^{5} D^{2} (2.58S_{60}+S_{100})
\end{equation}
where $L_{FIR}$ is the far-infrared luminosity in solar units, $D$ is the
distance to the galaxy in Mpc, $S_{60}$ is the $60\mu m$ flux in Jy and
$S_{100}$ is the $100\mu m$ flux in Jy.

Figure 13 shows a plot of the far-infrared luminosity normalised to K-band
luminosity against the overall EA to arm FWHM ratio. 
The correlation coefficient is 0.79 at a significance
level of 99.8\%. The error bars here are calculated from the errors on the
IRAS fluxes found using NED. IC 568 (denoted by a hollow circle) was found to
have a starburst nucleus on the basis of a very red nuclear colour (Paper
I), and this undoubtedly contributes to the high FIR luminosity. It was
omitted from the calculation of the correlation coefficient given above.

If the large--scale shock scenario (Roberts 1969) is correct a strong
correlation between arm strength (or the strength of shocks in the arms)
and star--formation rate would be expected. Knapen \& Beckman (1996) find
both star--formation rates and star formation efficiencies are enhanced in
the arms compared to the interarm regions of M101 and this supports the
Large Scale Shock Scenario. Figure 13 shows that we also find a good
correlation, in the expected sense.

\subsection{Comparison of B--band data with K-band data}

Additional B--band images were obtained for the galaxies NGC 5478, UGC 6958
and UGC 8939. The dominant source
of blue light is young bright stars and so these data can be used to
investigate the dependence of star formation on the underlying spiral
density perturbation. The B-band images were taken at the SAAO 1m
Elizabeth telescope with a 512x512 TEK8 CCD, and were reduced and
analysed in an essentially identical manner to that for the K-band data.
Here we compare the arm profiles, cross-sections and pitch angles derived
from K-band images with the same parameters from B-band data

Table 3 shows a comparison for the overall arm strengths between the
K--band data and B--band data. For the strongly armed galaxies 
(NGC 5478 and UGC 6958) the K--band arms are stronger than the B--band
arms. However, for UGC 8939 (which has weak arms) the B--band arms are 
stronger than the K--band arms. In our search for near neighbours we found
that UGC 8939 has two near neighbours at a distance of about 2 galaxy 
diameters and with similar surface brightnesses to UGC 8939. It may be
that the companion galaxies are affecting the star-formation rate in
UGC 8939, thus causing its B--band arms to appear stronger than its 
K--band arms.

If the large scale shock scenario (Roberts 1969) and the Lin-Shu
hypothesis (Lin \& Shu 1964, 1966) are correct we expect gas to approach
the arm from the concave side, assuming arms are trailing and that the
area under consideration is within corotation. The gas will then be
compressed and, if the perturbation is strong enough, shocked at the
centre of the arm, leading to subsequent star formation. Thus we should
see star formation on the leading, convex edges of spiral arms, within the
corotation radius.
Observationally, this should result in B-band arms leading K-band arms.
In order to test this we have compared the azimuthal angles at the start
of the arms, and the pitch angles, for the B-band and K-band images.
(Note that the image of UGC~6958 was first flipped to make the arms
spiral outwards in a clockwise direction, consistent with the other two
galaxies.) Table 4 shows the results of this analysis. Figure 14 also
shows a plot of the difference between the azimuthal angle of the K-band
arm and that of the B-band arm for UGC 8939 as radius increases.
It can be seen
that, generally, the B-band arms lead K-band arms by an average azimuthal
angle of $\simeq 5^{\circ}$ at low radii increasing to 10s of degrees at
high radii. This result and table 4 show that K-band arms are more tightly 
wound than B-band arms for UGC 8939. Similar plots for UGC 6958
and NGC 5478 (not shown), whilst noisier, show some tendency for K-band 
arms to have
smaller pitch angles than B--band arms. This confirms the result of Kennicutt
(1981) who used B-band and R-band optical data.
This therefore seems to
confirm the Large Scale Shock Scenario. However, there is another
possible explanation. Visible images show dust lying preferentially on
the trailing edges of arms, which would also result in B-band arms
appearing to lead K-band arms as a result of the increased extinction at
B. However, the fact that dust lanes appear on the trailing edges of arms
is in itself evidence for the Large Scale Shock Scenario (Roberts 1969)
since dust lanes are conventionally taken as tracers of shocks.

Figure 15 shows a comparison of B- and K-band arm cross-sections. In
general it can be seen that the FWHM of the cross-sections of B-band arms
are much smaller than those of K-band arms. There could be a number of
factors contributing towards this. In the B-band (or any visible
waveband) spiral arms are defined by the narrow regions where stars are
formed and this may be contributing to the low FWHM of B-band arms. In
addition, B-band arms are frequently observed to bifurcate into smaller,
narrower arms. Finally, blue light is strongly affected by dust which
lies predominantly on one side of spiral arms and would therefore have
the effect of making B-band arms appear narrower than K-band arms.

\section{Conclusions}

We now summarise the implications of these observations for the main
models of the formation and stability of spiral structure. No one theory
fits all of the observations, but a combination can explain most or all
of the features we have observed. This section summarises these
conclusions.

Simple density wave theories (e.g. Lin \& Shu 1964, 1966) are
undermined in three ways. Firstly, we do not get the correlation between
the fraction of light in the disk and pitch angle which is
predicted by Lin \& Shu (1964) and Roberts et al. (1975). Secondly, Lin
\& Shu (1964) predict that pitch angle should be the same for arms at
the same radius within a given galaxy. We find that this is not the case.
Finally, we have found that the FWHM of arm cross-sections are somewhat
narrower than predicted by simple density wave theories.

More complicated density wave theories, e.g. Modal theory (Bertin et al.
1989a, 1989b; Bertin \& Lin 1996), are supported by modulation
effects that we have found in arm strength as a function of radius, and
which appear to be fairly common.

Bar driven models of spiral structure (Sellwood \& Sparke 1988; Sanders
\& Huntley 1976) are not supported by our observations, as we find no
correlation between arm strength and bar strength, even in the inner part
of the disk.

Tidal effects from near neighbours seem to have some effect. We find that
4 out of 7 galaxies with arm EA greater than $35^{\circ}$ have near
neighbours whereas only 12 out of 45 galaxies in whole sample have near
neighbours. We also find that in the Fourier analysis of galaxies with
near neighbours, the even low-order modes are dominant, especially the
$m=2$ mode.

We have also discussed factors affecting star formation in the
disks of spiral galaxies.
We have found evidence both for and against the Large Scale
Shock Scenario (Roberts 1969). Firstly, we find a correlation
between arm contrast and the normalised star formation rate, 
in agreement with the predictions of the large scale shock scenario.
Secondly, we find that B-band arms are more loosely wound and 
narrower than K-band arms, confirming the predictions made
by density wave theories (Roberts 1969). However, this can also be
interpreted as an extinction effect because dust lies on the trailing
edges of arms.

Finally, and possibly most significantly, we find no correlation between
the arm properties of this sample of galaxies and their classified Hubble
type.  Combining this with the poor correlation between Hubble type and
K-band B/D ratio (Paper I), it would appear impossible to allocate a
Hubble type to these galaxies on the basis of these K-band observations, and
the morphology of the old stellar population varies surprisingly little
between Hubble types Sa and Sd. We speculate that the determining
parameter for Hubble type is cold gas content, which controls the star
formation rate and the distribution of the young stars, which dominate the
optical appearance.

\section{Acknowledgements}

We thank Paul Lynam for the B-band images, taken at the 1m Elizabeth
telescope at SAAO. We also wish to acknowledge Debra Elmegreen and
Frederick Chromey for useful suggestions, and John Porter for his
comments on an early draft of this paper. The United Kingdom Infrared 
Telescope is operated by the Joint Astronomy Centre on behalf of the 
U.K. Particle Physics and Astronomy Research Council.
This research has made use of the NASA/IPAC Extragalactic Database (NED),
which is operated by the Jet Propulsion Laboratory, California Institute of
Technology, under contract with the National Aeronautics and Space
Administration. We thank the referee for many helpful suggestions which
improved the content and presentation of this paper.

\renewcommand{\baselinestretch}{1}

\begin{table*}
\begin{center}
\caption{Summary of the arm and disk properties of the 45 galaxies.
The Hubble types of the galaxies are listed in column 2; 
the measured pitch angles of the spiral arms, averaged over the number 
of arms, with the radial extent over 
which this is measured in arcsec are listed in column 3; the overall arm
strengths, with the number of arms used to calculate this average in
brackets, in column 4; the dominant mode of the spiral structure in
column 5; the arm cross section full width at half maximum (FWHM) in
column 6; the fraction of light in the disk in column 7 and the 
azimuthal angle subtended by the strongest arm in column 8.}
\footnotesize
\begin{tabular}{l l c c c c c c}
\hline
Galaxy      & Hubble      & Pitch angle			             & Overall EA			& Dominant	& Arm 	 	& Fraction 	& Azimuthal angle	\\
Name	    & type        & 					     & (no. arms)			& Fourier       & FWHM		& of light	& subtended by arm	\\
	    &		  &					     &					& mode, m	&		& in disk	& in degrees		\\
\hline
ESO 555 -G 013 & Sa	  & $6.3^{\circ}\underline{+}0.7$ (7, 11)    & $19^{\circ}\underline{+}5$ (2)	& $2$		& $18^{\circ}$	& --		& 49			\\
IC 357	    & SBab	  & $8.3^{\circ}\underline{+}0.8$ (17, 23)   & $49^{\circ}\underline{+}7$ (2)	& $4$		& $15^{\circ}$  & 0.74		& 45			\\
IC 568	    & SBb	  & $8.4^{\circ}\underline{+}0.7$ (16, 24)   & $26^{\circ}\underline{+}4$ (2)	& $2$		& $18^{\circ}$  & 0.68		& 32			\\
IC 742	    & SBab	  & $9.3^{\circ}\underline{+}1.4$ (13, 17)   & $13^{\circ}\underline{+}3$ (2)	& $1$		& $20^{\circ}$  & 0.83		& 30			\\
IC 1196     & Scd	  & $15.8^{\circ}\underline{+}1.3$ (13, 21)  & $11^{\circ}\underline{+}2$ (2)	& $1$ 		& $18^{\circ}$  & 0.67		& 57			\\
IC 1764	    & SBb	  & $10.2^{\circ}\underline{+}0.6$ (15, 25)  & $18^{\circ}\underline{+}4$ (1)	& $2$		& $48^{\circ}$  & 0.78		& 31			\\
IC 1809	    & SBab	  & $9.9^{\circ}\underline{+}0.7$ (7, 20)    & $28^{\circ}\underline{+}2$ (2)	& $4$		& $18^{\circ}$  & 0.52		& 115			\\
IC 2363	    & SBbc	  & $9.8^{\circ}\underline{+}0.7$ (12, 20)   & $25^{\circ}\underline{+}5$ (3)	& $4$		& $28^{\circ}$  & 0.76		& 36			\\
IC 3692	    & SBa	  & $6.5^{\circ}\underline{+}1.0$ (15, 20)   & $11^{\circ}\underline{+}3$ (2)	& $1$		& $8^{\circ}$	& 0.70		& 88			\\
NGC 1219    & SAbc	  & $6.5^{\circ}\underline{+}2.0$ (8, 21)    & $9^{\circ}\underline{+}2$ (2)	& $4$		& $17^{\circ}$	& 0.84		& 89			\\
NGC 2416    & Scd	  & $9.1^{\circ}\underline{+}0.8$ (12, 22)   & $37^{\circ}\underline{+}6$ (2)	& $1$		& $17^{\circ}$	& 0.94		& 85			\\
NGC 2503    & SABbc	  & $6.1^{\circ}\underline{+}1.1$ (11, 19)   & $19^{\circ}\underline{+}2$ (2)	& $4$		& $26^{\circ}$	& 0.89		& 64			\\
NGC 2529    & SBd	  & $10.0^{\circ}\underline{+}0.5$ (10, 22)  & $31^{\circ}\underline{+}2$ (2)	& $1$		& $50^{\circ}$	& 0.9		& 83			\\
NGC 2628    & SABc	  & $9.4^{\circ}\underline{+}1.1$ (8, 27)    & $12^{\circ}\underline{+}4$ (3)	& $2$		& $30^{\circ}$	& 0.9		& 107			\\
NGC 3512    & SABc	  & $10.2^{\circ}\underline{+}0.9$ (11, 29)  & $12^{\circ}\underline{+}2$ (2)	& $1$		& $49^{\circ}$	& 0.92		& 173			\\
NGC 5478    & SABbc	  & $9.2^{\circ}\underline{+}1.2$ (6, 18)    & $37^{\circ}\underline{+}2$ (2)	& $1$		& $36^{\circ}$	& 0.73		& 93			\\
NGC 5737    & SBb	  & $9.4^{\circ}\underline{+}0.7$ (13, 22)   & $13^{\circ}\underline{+}1$ (2)	& $4$		& $32^{\circ}$	& 0.76		& 95			\\
NGC 6347    & SBb	  & $8.0^{\circ}\underline{+}0.8$ (5, 12)    & $11^{\circ}\underline{+}2$ (3)	& $4$		& $38^{\circ}$	& 0.93		& 200			\\
NGC 6379    & Scd	  & $8.7^{\circ}\underline{+}0.5$ (10, 23)   & $18^{\circ}\underline{+}7$ (2)	& $1$		& $24^{\circ}$	& 0.86		& 83			\\
NGC 6574    & SABbc	  & $9.9^{\circ}\underline{+}1.4$ (11, 28)   & $6^{\circ}\underline{+}3$ (3)	& $1$		& $20^{\circ}$	& 0.87		& 55			\\
UGC 850	    & SAbc	  & $4.8^{\circ}\underline{+}0.3$ (4, 16)    & $18^{\circ}\underline{+}1$ (2)	& $2$		& $30^{\circ}$	& 0.84		& 310			\\
UGC 1478    & SBc	  & $12.2^{\circ}\underline{+}0.5$ (8, 24)   & $39^{\circ}\underline{+}4$ (2)	& $1$		& $21^{\circ}$	& 0.74		& 54			\\
UGC 1546    & SABc	  & $8.0^{\circ}\underline{+}0.4$ (3, 15)    & $12^{\circ}\underline{+}2$ (1)	& $2$		& $17^{\circ}$	& 0.84		& 176			\\
UGC 2303    & SABb	  & $4.1^{\circ}\underline{+}0.5$ (8, 13)    & $9^{\circ}\underline{+}1$ (3)	& $3$		& $50^{\circ}$  & 0.74		& 125			\\
UGC 2585    & SBb	  & $4.6^{\circ}\underline{+}0.5$ (14, 17)   & $26^{\circ}\underline{+}1$ (1)	& $2$		& $22^{\circ}$	& 0.74		& 54			\\
UGC 2705    & SBcd	  & $7.7^{\circ}\underline{+}1.0$ (4, 24)    & $14^{\circ}\underline{+}2$ (3)	& $3$		& $44^{\circ}$	& 0.66		& 184			\\
UGC 2862    & SABa	  & $6.9^{\circ}\underline{+}0.4$ (5, 13)    & $10^{\circ}\underline{+}2$ (1)	& $3$		& $38^{\circ}$	& 0.34		& 76			\\
UGC 3053    & Scd	  & $8.6^{\circ}\underline{+}0.3$ (7, 16)    & $8^{\circ}\underline{+}2$ (3)	& $1$		& $28^{\circ}$	& 0.94		& 114			\\
UGC 3091    & SABd	  & $9.1^{\circ}\underline{+}0.9$ (5, 10)    & $8^{\circ}\underline{+}2$ (2)	& $1$		& $43^{\circ}$	& 0.87		& 101			\\
UGC 3171    & SBcd	  & $12.9^{\circ}\underline{+}0.4$ (5, 11)   & $10^{\circ}\underline{+}3$ (2)	& $1$		& $32^{\circ}$  & 0.96		& 190			\\
UGC 3233    & Scd	  & $9.0^{\circ}\underline{+}1.6$ (7, 10)    & $5^{\circ}\underline{+}1$ (2)	& $2$		& $47^{\circ}$	& 0.75		& 69			\\
UGC 3296    & Sab	  & $10.5^{\circ}\underline{+}1.4$ (14, 20)  & $6^{\circ}\underline{+}1$ (3)	& $2$		& $33^{\circ}$	& 0.88		& 63			\\
UGC 3578    & SBab	  & $9.3^{\circ}\underline{+}0.6$ (15, 32)   & $24^{\circ}\underline{+}9$ (2)	& $2$		& $19^{\circ}$	& 0.74		& 71			\\
UGC 3707    & Sab	  & $7.0^{\circ}\underline{+}1.4$ (15, 19)   & $33^{\circ}\underline{+}4$ (2)	& $1$		& $22^{\circ}$	& 0.73		& 17			\\
UGC 3806    & SBcd	  & $5.8^{\circ}\underline{+}0.3$ (8, 30)    & $8^{\circ}\underline{+}2$ (2)	& $1$		& $26^{\circ}$	& 0.93		& 72			\\
UGC 3839    & SBb	  & $8.7^{\circ}\underline{+}0.8$ (13, 18)   & $28^{\circ}\underline{+}2$ (2)	& $3$		& $21^{\circ}$  & 0.80		& 39			\\
UGC 3900    & SBb	  & $7.1^{\circ}\underline{+}0.2$ (13, 25)   & $10^{\circ}\underline{+}3$ (2)	& $1$		& $26^{\circ}$	& 0.78		& 154			\\
UGC 3936    & SBbc	  & $7.6^{\circ}\underline{+}0.3$ (10, 27)   & $9^{\circ}\underline{+}4$ (2)	& $4$		& $24^{\circ}$  & 0.81		& 76			\\
UGC 4643    & SAbc	  & $4.1^{\circ}\underline{+}0.6$ (9, 30)    & $31^{\circ}\underline{+}2$ (3)	& $4$		& $48^{\circ}$	& 0.72		& 214			\\
UGC 5434    & SABb	  & $8.7^{\circ}\underline{+}0.5$ (8, 15)    & $37^{\circ}\underline{+}5$ (2)	& $2$		& $31^{\circ}$	& 0.93		& 130			\\
UGC 6166    & Sbc	  & $10.6^{\circ}\underline{+}1.0$ (10, 22)  & $6^{\circ}\underline{+}1$ (1)	& $2$		& $20^{\circ}$	& 0.89		& 45			\\
UGC 6332    & SBa	  & $7.6^{\circ}\underline{+}0.2$ (14, 27)   & $24^{\circ}\underline{+}4$ (2)	& $4$		& $37^{\circ}$	& 0.67		& 51			\\
UGC 6958    & SABbc	  & $7.4^{\circ}\underline{+}0.4$ (8, 16)    & $38^{\circ}\underline{+}2$ (1)	& $2$		& $72^{\circ}$	& 0.89		& 44			\\
UGC 8939    & SABb	  & $5.3^{\circ}\underline{+}0.6$ (7, 15)    & $10^{\circ}\underline{+}2$ (2)	& $4$		& $15^{\circ}$	& 0.72		& 128			\\
UGC 11524   & SAc	  & $7.5^{\circ}\underline{+}0.5$ (8, 27)    & $20^{\circ}\underline{+}4$ (2)	& $2$		& $63^{\circ}$	& 0.82		& 152			\\
\hline
\end{tabular}
\normalsize
\label{armprop}
\end{center}
\end{table*}

\begin{table*}
\begin{center}
\caption{Relative Strengths of the Fourier modes}
\footnotesize
\begin{tabular}{l c c c c}
\hline
	& \multicolumn{4}{c}{Fourier Mode} \\
\hline
Galaxy Name	& m=1		& m=2		& m=3		& m=4		\\
\hline
ESO 555 -G 013	& 1.9		& 2.5		& 4.2		& 1.8		\\
IC 357		& 1.7		& 0.6		& 1.0		& 1.9		\\
IC 568		& 1.0		& 2.1		& 0.7		& 1.0		\\
IC 742		& 4.8		& 0.3		& 0.8		& 2.4		\\
IC 1196		& 2.6		& 0.3		& 1.6		& 1.1		\\
IC 1764		& 1.9		& 2.2		& 0.9		& 2.0		\\
IC 1809		& 1.0		& 0.7		& 1.3		& 2.1		\\
IC 2363		& 0.1		& 1.2		& 2.6		& 6.4		\\
IC 3692		& 1.3		& 1.0		& 1.0		& 0.9		\\
NGC 1219	& 2.1		& 0.4		& 1.5		& 2.8		\\
NGC 2416	& 1.9		& 0.9		& 0.3		& 0.6		\\
NGC 2503	& 1.2		& 1.8		& 0.3		& 2.2		\\
NGC 2529	& 1.5		& 0.6		& 1.4		& 1.3		\\
NGC 2628	& 1.0		& 3.0		& 0.5		& 0.9		\\
NGC 3512	& 2.2		& 1.1		& 1.6		& 1.1		\\
NGC 5478	& 1.6		& 1.2		& 1.5		& 1.1		\\
NGC 5737	& 1.7		& 0.7		& 0.7		& 2.4		\\
NGC 6347	& 0.9		& 2.3		& 0.2		& 3.8		\\
NGC 6379	& 3.1		& 0.9		& 0.3		& 0.9		\\
NGC 6574	& 3.6		& 0.4		& 1.3		& 3.5		\\
UGC 850		& 0.2		& 4.1		& 1.5		& 2.2		\\
UGC 1478	& 5.7		& 2.8		& 1.5		& 1.2		\\
UGC 1546	& 1.8		& 7.8		& 0.4		& 1.3		\\
UGC 2303	& 1.4		& 1.2		& 2.9		& 1.9		\\
UGC 2585	& 2.1		& 4.7		& 1.3		& 0.4		\\
UGC 2705	& 1.1		& 2.2		& 3.7		& 0.9		\\
UGC 2862	& 0.9		& 1.0		& 1.1		& 0.9		\\
UGC 3053	& 2.1		& 1.3		& 1.8		& 0.2		\\
UGC 3091	& 2.8		& 1.6		& 2.2		& 2.3		\\
UGC 3171	& 1.9		& 0.7		& 0.7		& 0.8		\\
UGC 3233	& 2.6		& 3.0		& 0.7		& 1.8		\\
UGC 3296	& 1.1		& 4.8		& 2.6		& 1.0		\\
UGC 3578	& 1.0		& 1.2		& 0.9		& 0.9		\\
UGC 3707	& 7.0		& 1.5		& 1.4		& 0.2		\\
UGC 3806	& 1.5		& 0.7		& 0.9		& 1.5		\\
UGC 3839	& 1.0		& 1.1		& 1.9		& 1.6		\\
UGC 3900	& 3.5		& 0.9		& 0.3		& 0.5		\\
UGC 3936	& 1.1		& 0.7		& 1.6		& 2.2		\\
UGC 4643	& 0.2		& 1.9		& 0.7		& 1.9		\\
UGC 5434	& 0.4		& 3.7		& 0.7		& 3.4		\\
UGC 6166	& 0.5		& 1.1		& 0.1		& 0.8		\\
UGC 6332	& 2.2		& 0.6		& 0.3		& 4.0		\\
UGC 6958	& 0.9		& 8.7		& 1.2		& 5.9		\\
UGC 8939	& 1.3		& 0.9		& 0.9		& 2.0		\\
UGC 11524	& 1.3		& 1.8		& 1.6		& 0.8		\\
\hline
\end{tabular}
\normalsize
\end{center}
\end{table*}

\begin{table*}
\begin{center}
\caption{Comparison of overall K--band arm strengths with overall B--band arm strengths. Arm strengths are measured in terms of EA in degrees.}
\footnotesize
\begin{tabular}{l c c}
\hline
Galaxy		& \multicolumn{2}{c}{Overall Arm strength}	\\
\hline
		& K--band		& B--band		\\
\hline
NGC 5478	& 37.0$\pm$1.9		& 19.5$\pm$2.0		\\
UGC 6958	& 38.0$\pm$2.4		& 15.4$\pm$1.3		\\
UGC 8939	& 10.4$\pm$2.1		& 23.4$\pm$0.7		\\
\hline
\end{tabular}
\normalsize
\end{center}
\end{table*}

\begin{table*}
\begin{center}
\caption{Comparison of arm properties between B-band and K-band images measured for the strongest arm in the galaxies.}
\footnotesize
\begin{tabular}{l c c c c}
\hline
Galaxy &
\multicolumn{2}{c}{B-band} &
\multicolumn{2}{c}{K-band} \\
\hline
	& Pitch Angle	& Azimuthal angle & Pitch Angle & Azimuthal angle \\
	&		& at start of arm & 		& at start of arm \\
\hline
NGC 5478 & $10.9\underline{+}1.3$ & $224^{\circ}$ & $9.9\underline{+}1.1$ & $219^{\circ}$ \\
UGC 6958 & $9.3\underline{+}1.5$ & $347^{\circ}$ & $7.4\underline{+}1.4$ & $342^{\circ}$ \\
UGC 8939 & $5.5\underline{+}0.6$ & $363^{\circ}$ & $5.1\underline{+}0.6$ & $355^{\circ}$ \\
\hline
\end{tabular}
\normalsize
\end{center}
\label{azangle}
\end{table*}

\clearpage

\begin{flushleft}

Figure 1: Arm profiles in terms of arm equivalent angle versus radius.

Figure 2: Arm cross-sections (top), symmetric components (central) and 
antisymmetric components 
(lower). Leading edges (i.e. the convex sides) of arms are plotted on the 
right. A 540$^{\circ}$ cut of IC 1809 is also shown (bottom).

Figure 3: The occurrence of the strongest Fourier modes
found in the disks of the entire sample compared with 
(a) Galaxies with companions within 6 galaxy diameters shaded in black;
(b) Galaxies with bars stronger than the average overall bar strength shaded
in black.

Figure 4: Pitch angle versus fraction of K-band light in the disk.

Figure 5: Pitch angle versus Hubble type. The line represents the
correlation predicted by Roberts et al. (1975).

Figure 6: The galaxy NGC 2503 in polar coordinates. $100\ln{(r)}$ is plotted 
in the x direction where $r$ is measured in pixels. $\theta$ is plotted in the 
y direction in units of degrees, so the plot represents a turn and a half
around the galaxy. 
The arms appear straight (with a constant gradient) in 
this plot and their pitch angles can be measured directly from their gradients.
NGC 2503 demonstrates that pitch angle is not the same even for arms within
the same galaxy.

Figure 7: Arm strength in the inner part of the disk versus bar strength for
all the barred galaxies in the sample. Bar strength and arm strength are both 
measured in terms of EA in degrees.

Figure 8: Greyscale unsubtracted and difference images 
of the galaxy UGC 3900 showing that its 
arms start ahead of the end of its bar and that its bulge is offset 
from the centre of the disk. The axes are labelled in units of
pixels, where 1 pixel = 0.286 arcsec.

Figure 9: Arm strengths in terms of EA; galaxies with near neighbours
are shaded in black.

Figure 10: Distribution of the azimuthal angles subtended by arms;
galaxies with near neighbours are shaded in black.

Figure 11: Azimuthal angle subtended by arms (in degrees)
against bulge--to--disk ratio.

Figure 12: Arm strength (in terms of EA in degrees) versus 
disk central surface 
brightness (K-mag per square arcsec).

Figure 13: Far-infrared bolometric luminosity normalised to K-band magnitude
versus arm contrast. The point 
represented by a hollow circle is IC 568, 
which we believe has a starburst nucleus that probably dominates its FIR 
luminosity. The line is an error weighted regression, ignoring IC 568.

Figure 14: Radial dependence of the difference in azimuthal angle between
B--band and K--band arms.

Figure 15: Comparison of K-band arm cross-sectional profiles with B-band
cross-sectional profiles. K-band data are represented by circles and
B-band data are represented by squares.

\end{flushleft}

\clearpage

\pagestyle{empty}

\begin{figure}
\includegraphics{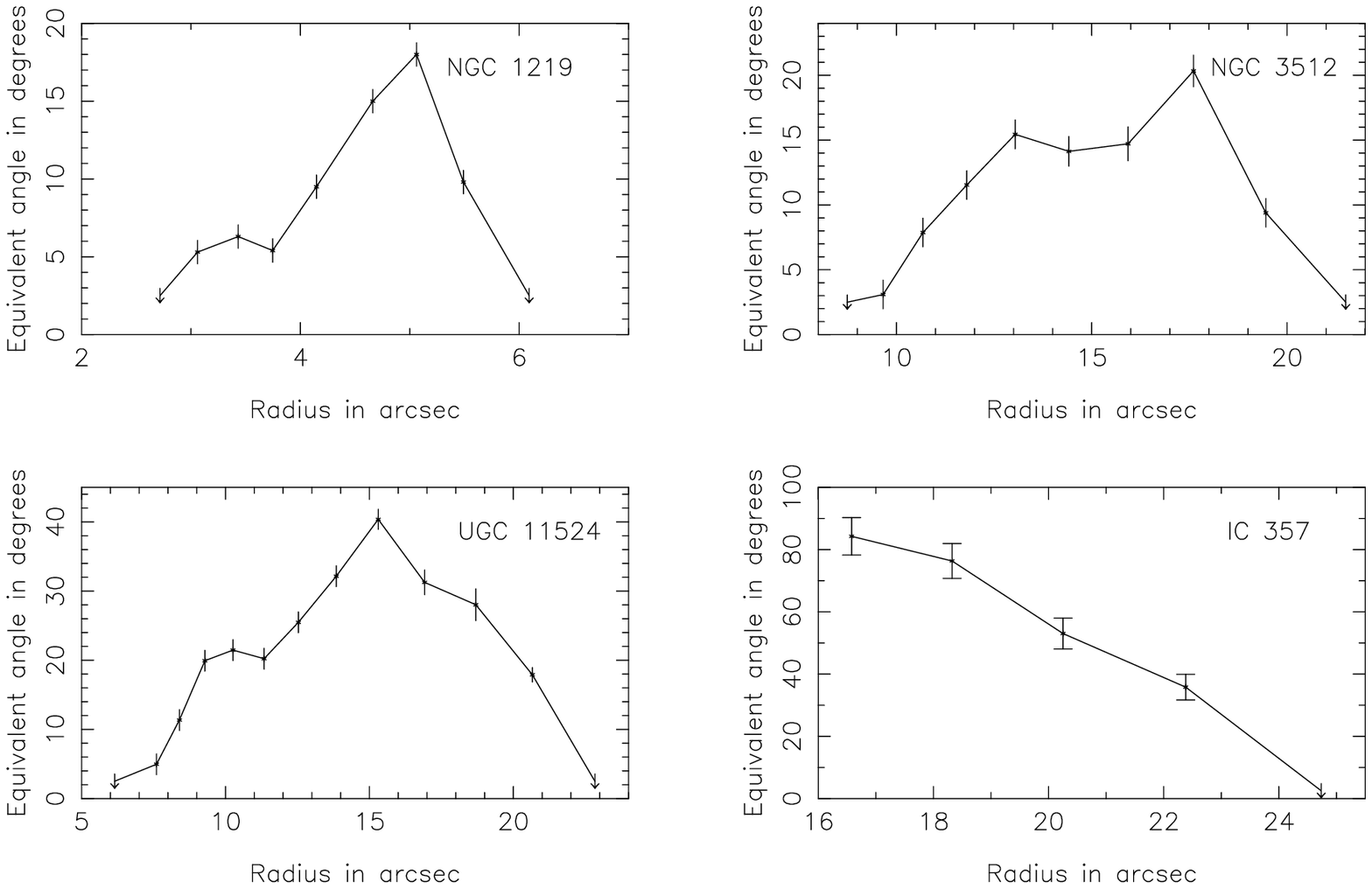}
\vspace*{20cm}
\end{figure}

\clearpage

\begin{figure}
\includegraphics{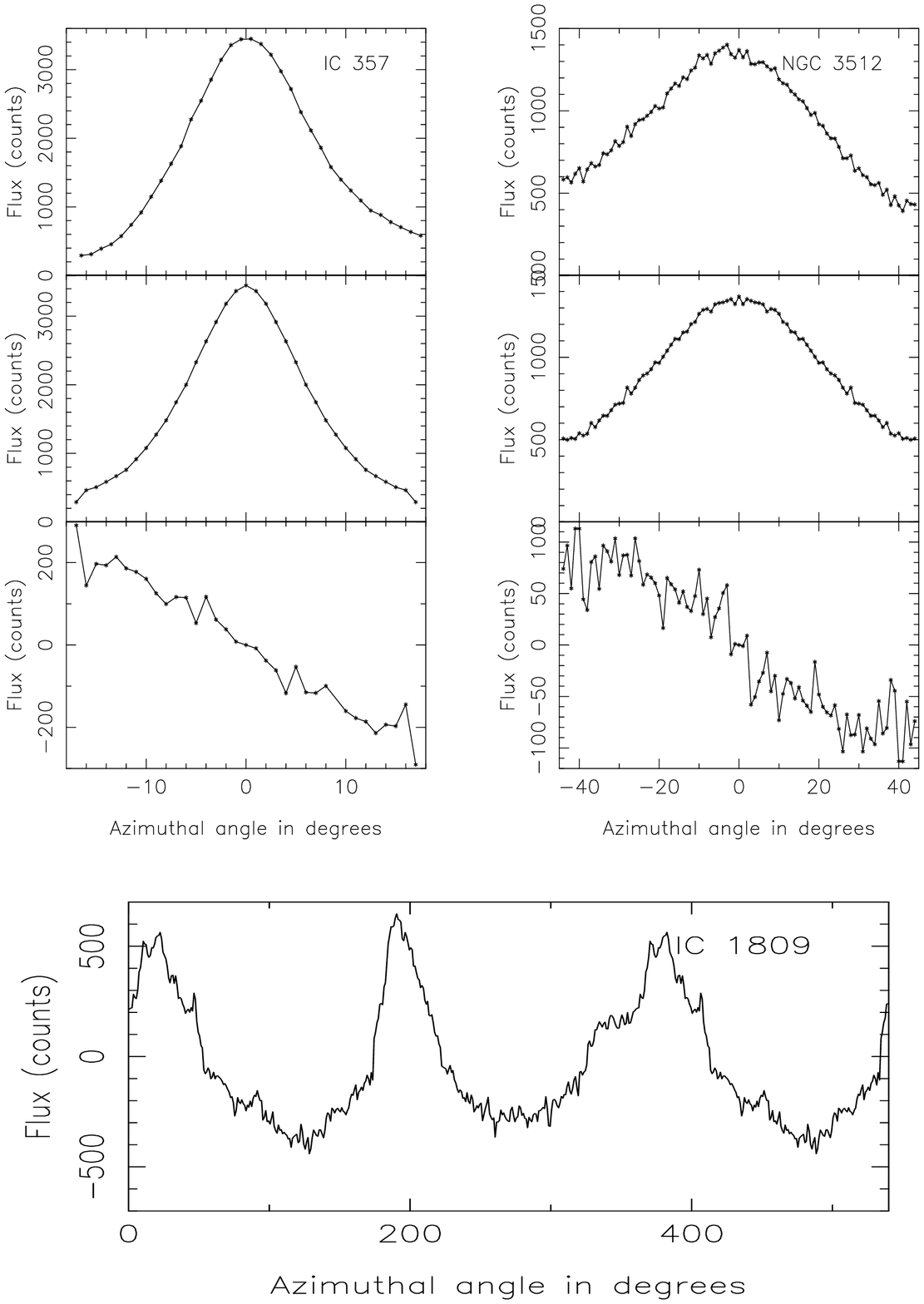}
\vspace*{20.0cm}
\end{figure}

\clearpage

\begin{figure}
\includegraphics{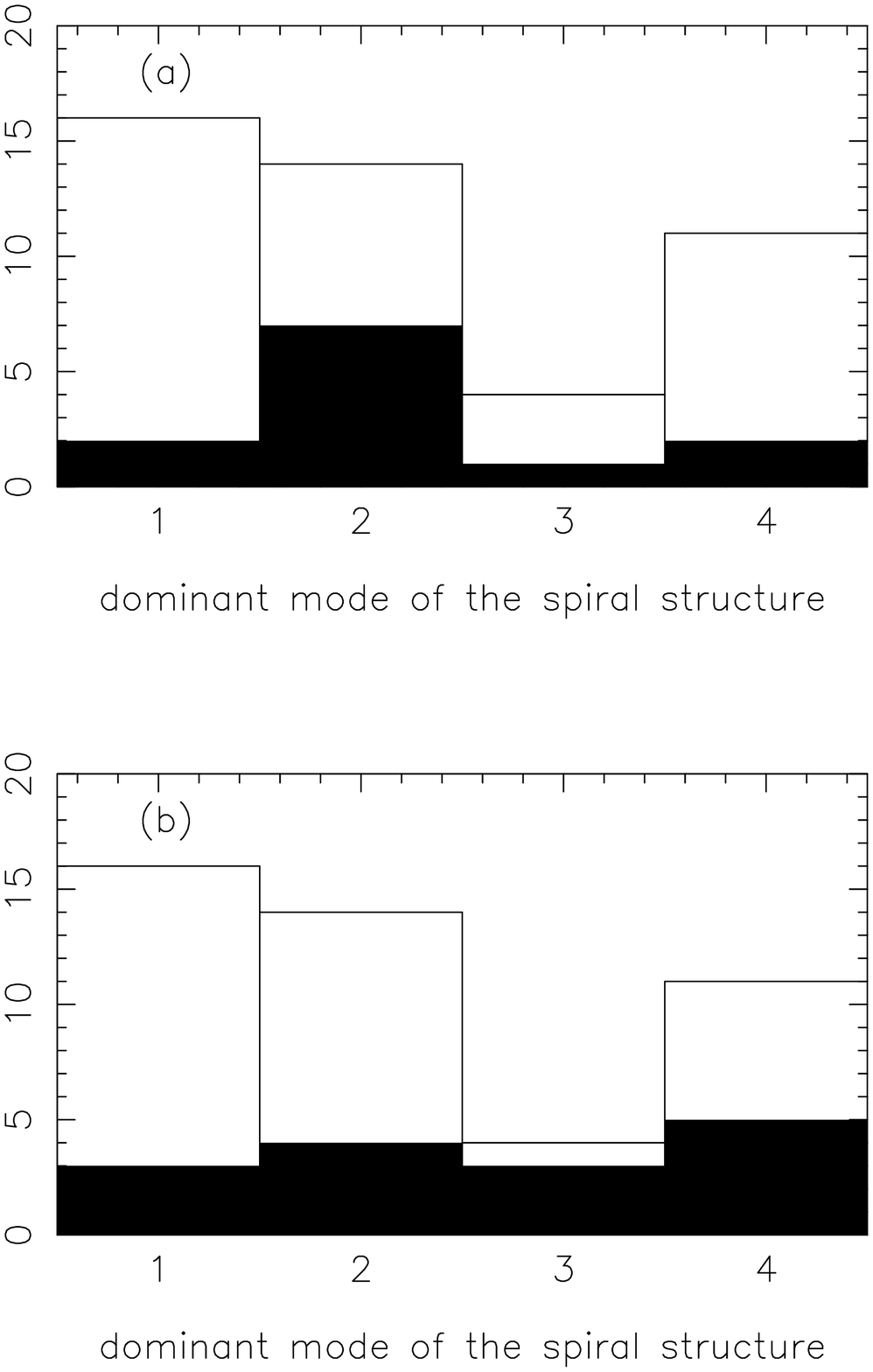}
\vspace*{15cm}
\end{figure}

\clearpage

\begin{figure}
\includegraphics{figure4.ps}
\vspace*{15cm}
\end{figure}

\clearpage

\begin{figure}
\includegraphics{figure5.ps}
\vspace*{15cm}
\end{figure}

\clearpage

\begin{figure}
\includegraphics{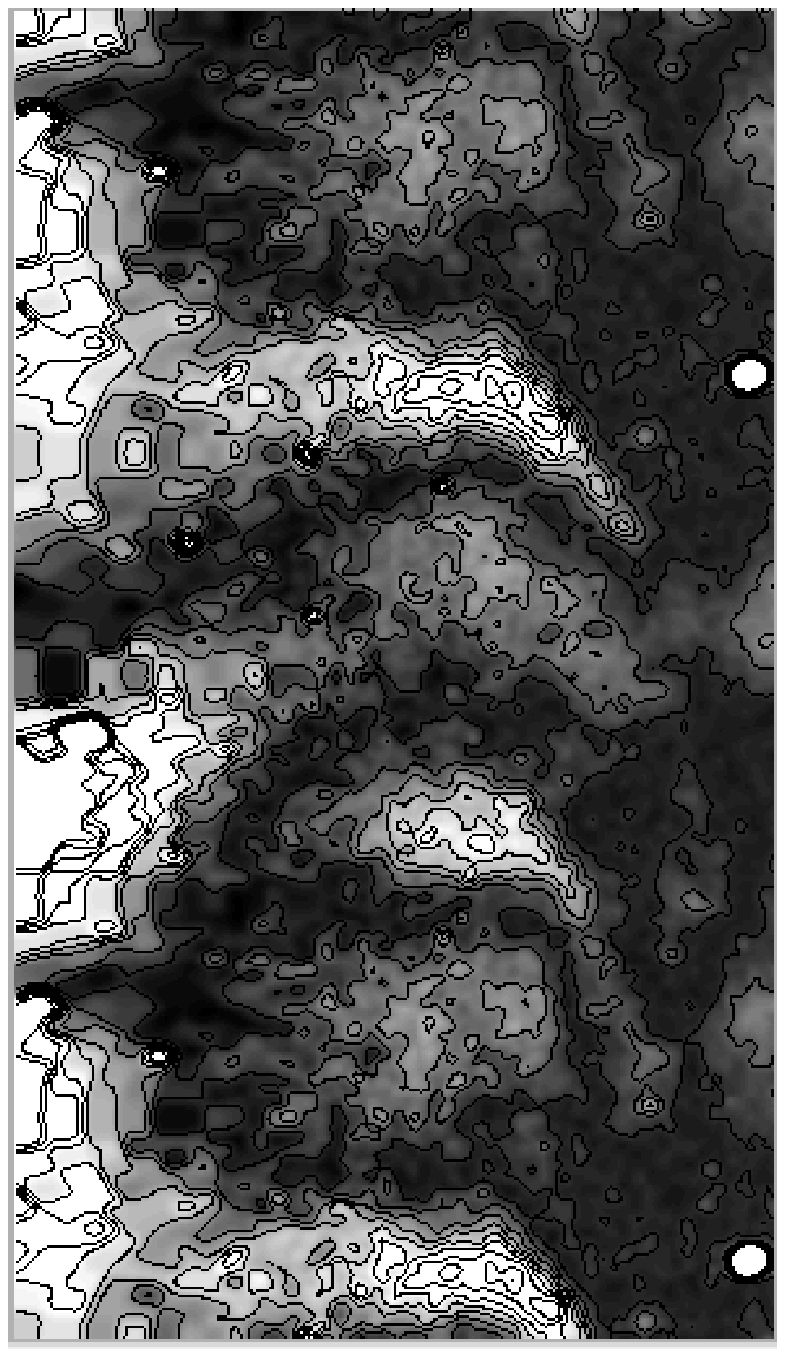}
\vspace*{15cm}
\end{figure}

\clearpage

\begin{figure}
\includegraphics{figure7.ps}
\vspace*{15cm}
\end{figure}

\clearpage

\begin{figure}
\includegraphics{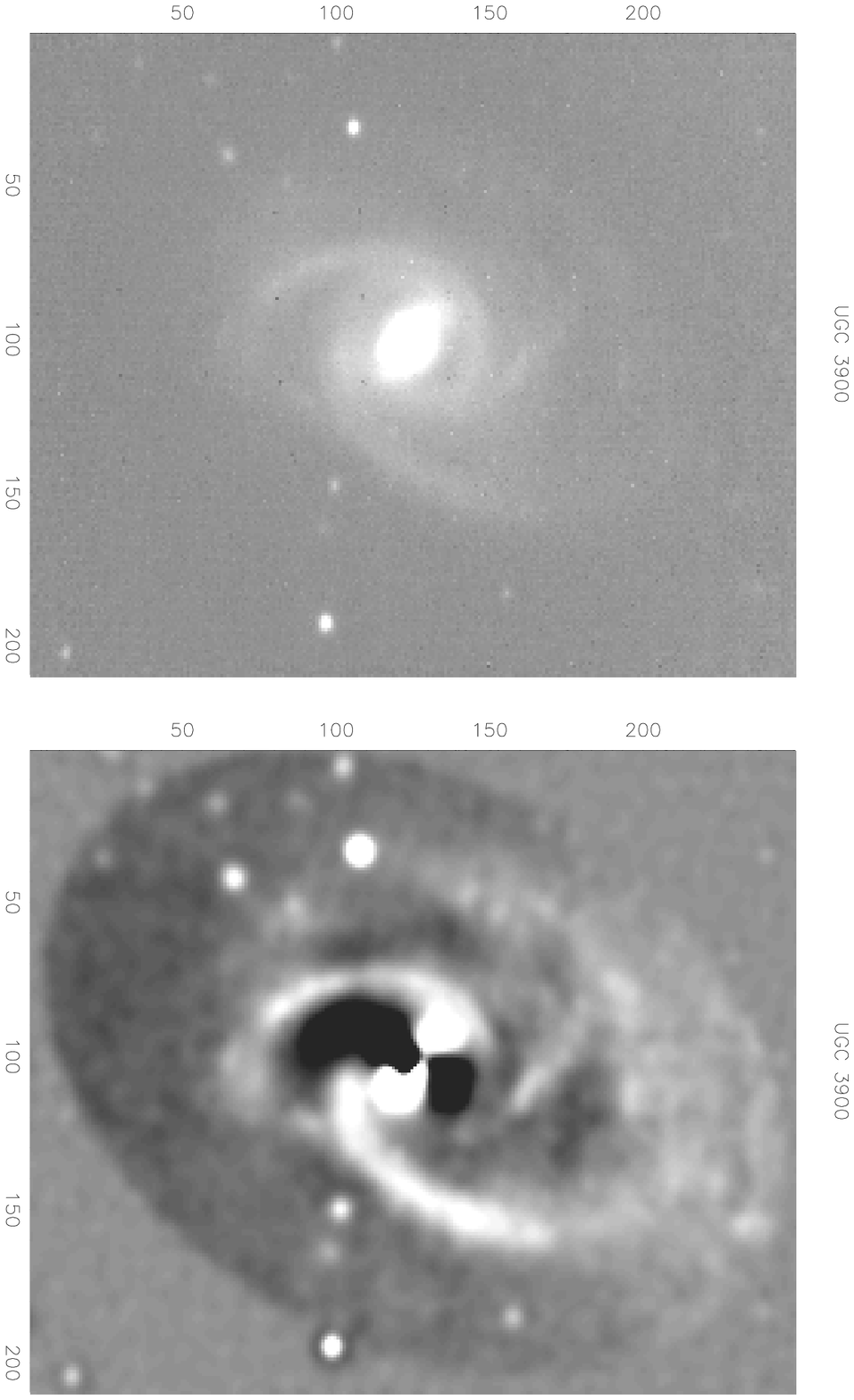}
\vspace*{17cm}
\end{figure}

\clearpage

\begin{figure}
\includegraphics{figure9.ps}
\vspace*{15cm}
\end{figure}

\clearpage

\begin{figure}
\includegraphics{figure10.ps}
\vspace*{15cm}
\end{figure}

\clearpage

\begin{figure}
\includegraphics{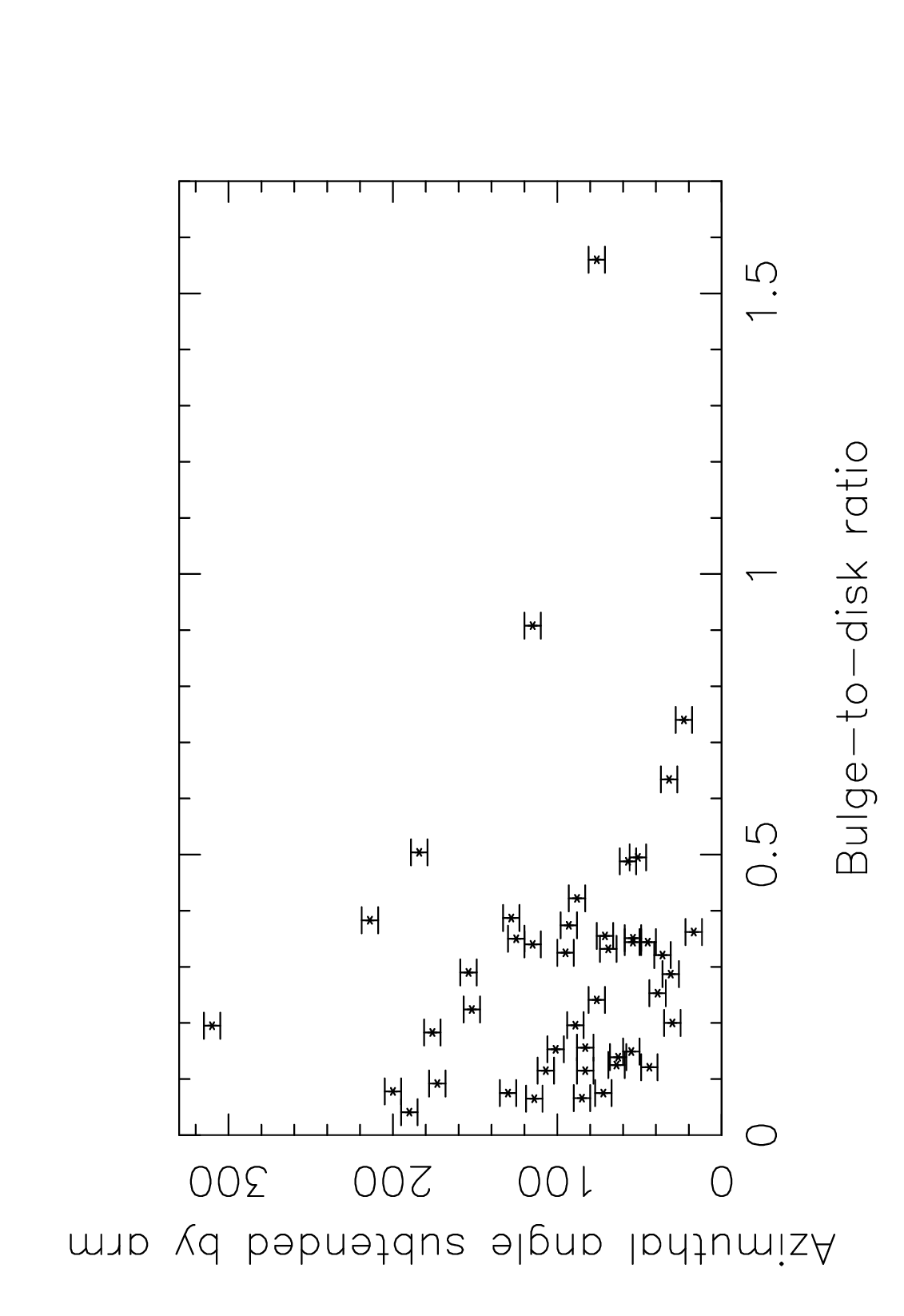}
\vspace*{15cm}
\end{figure}

\clearpage

\begin{figure}
\includegraphics{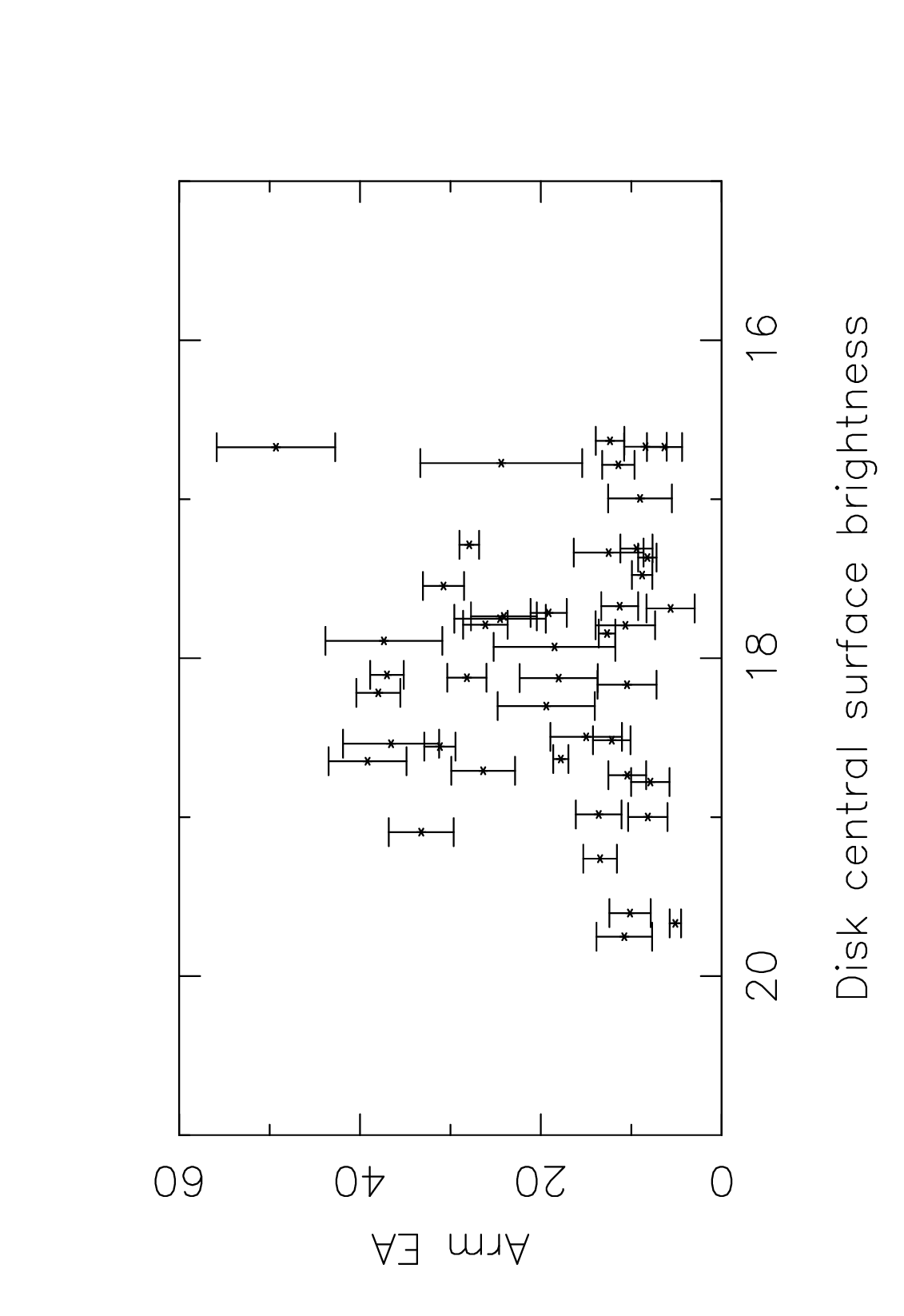}
\vspace*{15cm}
\end{figure}

\clearpage

\begin{figure}
\includegraphics{figure13.ps}
\vspace*{15cm}
\end{figure}

\clearpage

\begin{figure}
\includegraphics{figure14.ps}
\vspace*{15cm}
\end{figure}

\clearpage

\begin{figure}
\includegraphics{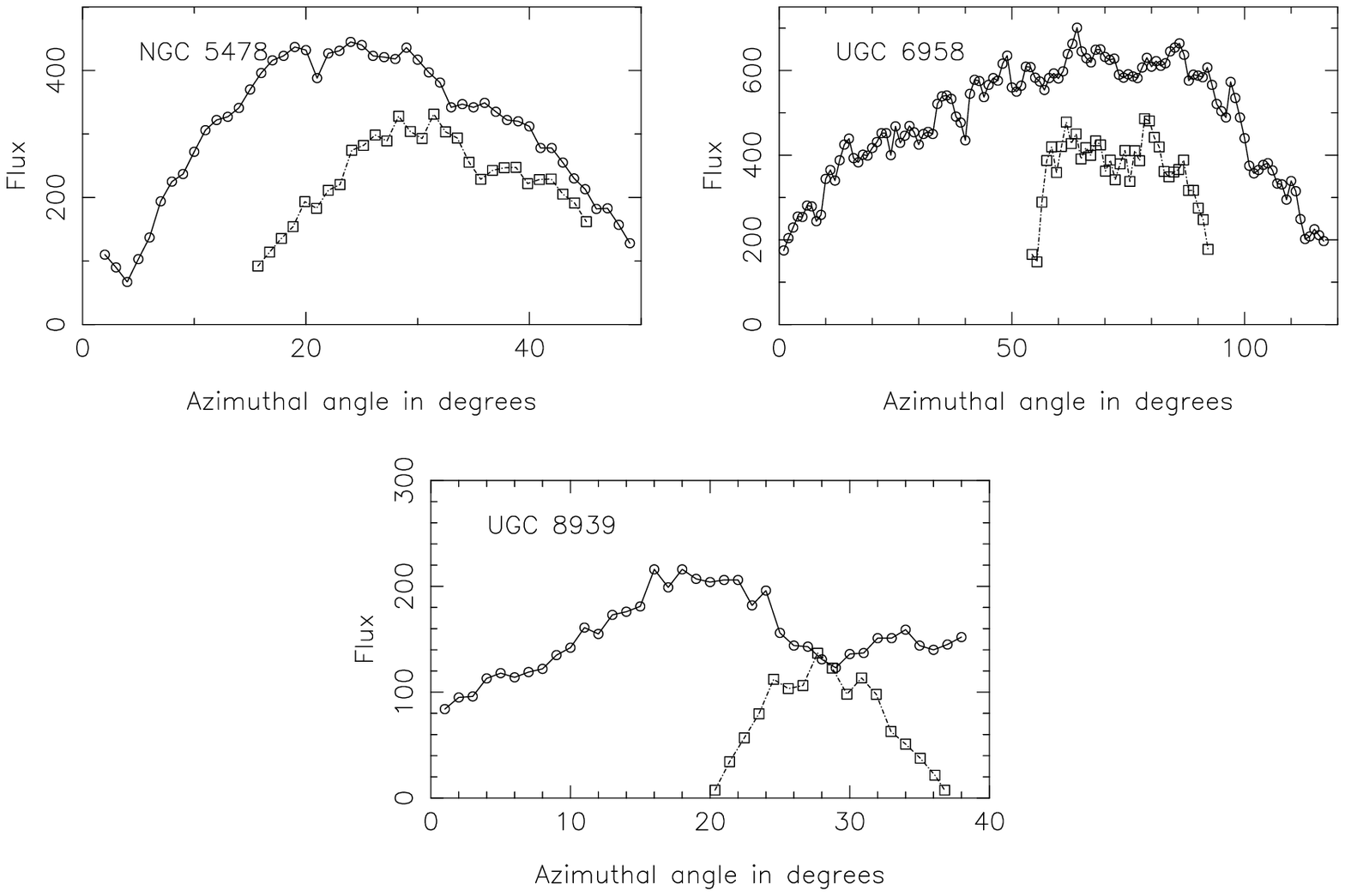}
\vspace*{20cm}
\end{figure}

\clearpage

\end{document}